\documentclass[twocolumn]{webofc}
\usepackage[varg]{txfonts}   
\begin{document}
\title{CAMEA ESS - The Continuous Angle Multi-Energy Analysis Indirect Geometry Spectrometer for the European Spallation Source.}
%
%

\author{P. G. Freeman \inst{1}\fnsep\thanks{\email{paul.freeman@epfl.ch}}
            \and
        J. O. Birk\inst{2}
             \and
       M. Mark\'{o}\inst{3}
             \and
     M. Bertelsen\inst{2}
         \and
        J. Larsen\inst{4}
             \and
        N. B. Christensen\inst{4}
             \and
        K. Lefmann\inst{2}
          \and
        J. Jacobsen\inst{2}
         \and
             Ch. Niedermayer\inst{3}
          \and
             F. Juranyi\inst{3}
          \and
        H. M. Ronnow \inst{1}\fnsep\thanks{\email{henrik.ronnow@epfl.ch}}
             }

\institute{Laboratory for Quantum Magnetism, ICMP,
\'{E}cole Polytechnique F\'{e}d\'{e}rale de Lausanne (EPFL), CH-1015 Lausanne, Switzerland
\and
        Niels Bohr Institute, University of Copenhagen, Universitetsparken 5, 2100 Copenhagen, Denmark.
\and
           Laboratory of Neutron Scattering, Paul Scherrer Institute, 5232 Villigen–PSI , Switzerland.
\and
            Department of Physics, Technical University of Denmark, Kongens Lyngby 2800, Denmark.
}

\abstract{ The CAMEA ESS neutron spectrometer is designed to achieve a high detection efficiency in the horizontal scattering plane, and to maximize the use of the long pulse European Spallation Source. It is an indirect geometry time-of-flight spectrometer that uses crystal analysers to determine the final energy of neutrons scattered from the sample. Unlike other indirect gemeotry spectrometers CAMEA will use ten concentric arcs of analysers to analyse scattered neutrons at ten different final energies, which can be increased to 30 final energies by use of prismatic analysis. In this report we will outline the CAMEA instrument concept, the large performance gain,  and the potential scientific advancements that can be made with this instrument.
}
\maketitle
\section{Introduction}
\label{intro}

For measuring excitations such as phonons and magnons in materials there are two dominant types of inelastic neutron spectrometers  presently in use, i) direct geometry time-of-flight (ToF) spectrometers and ii) triple-axis spectrometers (TAS). Direct ToF instruments compensate a low incident neutron flux from using monochromatic pulses, by detecting the scattered neutrons of all energies over a large solid angle in position sensitive detectors. In contrast TAS focus a continuous high flux monochromatic beam of neutrons on a sample but have a low detection coverage of counting neutrons at one final neutron energy at one position, or tens of angles in the case of multiplexed TAS\cite{multiplex,MACS,flatcone}.

The European Spallatioon Source (ESS) will be a 5\,MW long pulse spallation neutron source, and will have the world's highest peak brightness for cold neutrons\cite{Peggs2013}. The time averaged cold neutron flux of the ESS will also be greater than that of world leading continuous sources such as the Insitut Laue-Langevin high flux reactor. The instrument design phase for the ESS is an ideal opportunity to consider new possibilities for instrument concepts.


Researchers of magnetism are the largest user community of single crystal spectrometers. This community often use applied magnetic fields to tune the magnetic properties of materials across phase transitions in to new phases of matter, where inelastic neutron scattering uniquely determines the nature of the magnetic phase.
Cryomagnets used in these experiments however restrict the access for neutrons. For example, currently the
highest vertical field cryomagnetic for neutron spectroscopy 
is the 16 T `Fat Sam' produced by Bruker, for the Spallation Neutron Source (SNS) as a collaborative project between the Swiss Neutron Scattering Society (SGN/SSDN) and the Oak Ridge National Laboratory\cite{fatsam}. This magnet has a vertical opening angle of $\pm 4^{\circ}$. When used on the direct geometry ToF Cold Chopper Neutron Spectrometer (CNCS) at SNS, 75\% of CNCS's detectors out of the horizontal plane are blocked, reducing the instruments efficiency. TAS spectrometers operate in the horizontal plane and are therefore less restricted by the neutron access of split-pair cryomagnets. A similar situation holds for using anvil cells for extreme pressure to tune magnetic properties of materials\cite{anvil}. TAS spectrometers are the instrument of choice for these types of experiments, however a TAS would not take advantage of the pulsed nature of the ESS. This provides an initial motivation to examine optimizing a spallation source instrument that maximizes the neutron count rate within the horizontal plane.

\section{Concept}
\label{concept}

\begin{figure}
\includegraphics[width=7.5cm,clip]{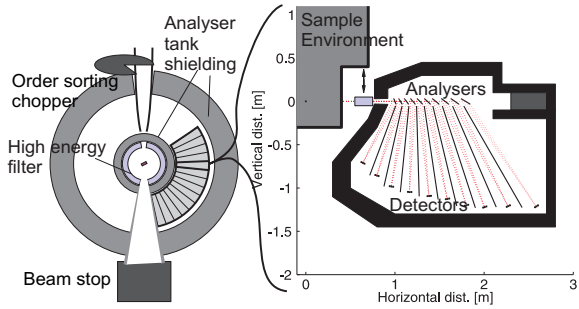}
\caption{Left:A simplified overview of the CAMEA spectrometer from the end of the neutron guide, not to scale.  The sample is surrounded on one side by the analyser-detector chamber that covers a large angle within the horizontal
plane. A cross section to scale of one 9$^{\circ}$ multi-analyser-detector module is shown on the right. There is
large 90 cm diameter space for sample environment, and a removable cooled Be high energy filter placed in front of
the entry to the secondary spectrometer. We show how 10 analysers sat behind each other working at different final
 neutron energies can geometrically be spaced with neutron shielding between the analyser to detector channels.
 After the last analyser the neutrons are directed into a get lost tube into a beamstop.}
\label{Overview}       
\end{figure}

We directed our attention to indirect geometry ToF spectrometry where the final energy of the neutron is determined by a crystal analyser.  Present Indirect spectrometers analyse neutrons scattered from a sample at one final fixed energy, from knowing the final energy and the time-of-flight the scattering process is determined. Present indirect ToF spectrometers are inefficient as the scattered neutron energy is only analysed once, any neutron that does not have the correct final energy only increases  the background signal. We note that the indirect spectrometers such as PRISMA (ISIS)\cite{PRISMA} and CQS (Los Alamos)\cite{CQS} did work with variable final neutron energies, but only analysed a scattered neutron's energy a single time. The majority of spallation source indirect ToF spectrometers are ultra high resolution backscattering spectrometers, or vibrational spectrometers that measure phonon density of states. Neither of these two instrument classes are ideally suited for mapping phonon or magnon dispersion curves in single crystals, however the back scattering spectrometer Osiris at the ISIS facility is successfully used to study magnetic excitations in single crystals\cite{osiris}.

The essential evolution in neutron instrumentation of CAMEA is the secondary spectrometer:
\begin{description} \itemsep1pt \parskip0pt \parsep0pt
  \item[1)] Vertically scattering analysers that allows for increased coverage of in plane scattering.
  \item[2)] Multiple concentric arcs of analysers sat behind each other to perform multiple final energy analysis.
  \item[3)] Use of Position Sensitive Detectors (PSDs) for quasi-continuous angular coverage.
   \item[4)] Prismatic analysis from a distance collimated analyser allowing multi-energy analysis from a single analyser.
   \item[5)] An order sorting chopper that enables use of first and second order reflections off the analysers.
\end{description}


Neutrons have large penetration depths, for  analyser crystals of pyrolytic graphite (PG) of 1\,mm thickness mounted on 1\,mm Si wafers, the transmission rate we have experimentally determined as $>98\,\%$. Typically 2\,mm PG analyser crystals are used, but reducing to 1\,mm halves the cost, increases transmission, for only a small cost in reflectivity. We propose an instrument with 10 concentric arcs of analysers that direct analysed neutrons vertically into detectors, if a neutron scattered form the sample is not at the energy of the first analyser the neutron is transmitted to be analysed by up to nine further analysers working at different final energies.   Scattering vertically  has been shown not to reduce energy resolution when studying samples of small vertical height (<1\,cm)\cite{flatcone}. The increased efficiency of this Multi-Energy Analysis for 10 energies gives a gain factor $>9.1$, considering transmission rates. In figure \ref{Overview}(a) we sketch an overview of the secondary spectrometer, and in Fig.\ref{Overview}(b) we outline how ten analysers placed behind each other can be positioned.

\begin{figure*}
\centerline{\includegraphics[width=14cm,clip]{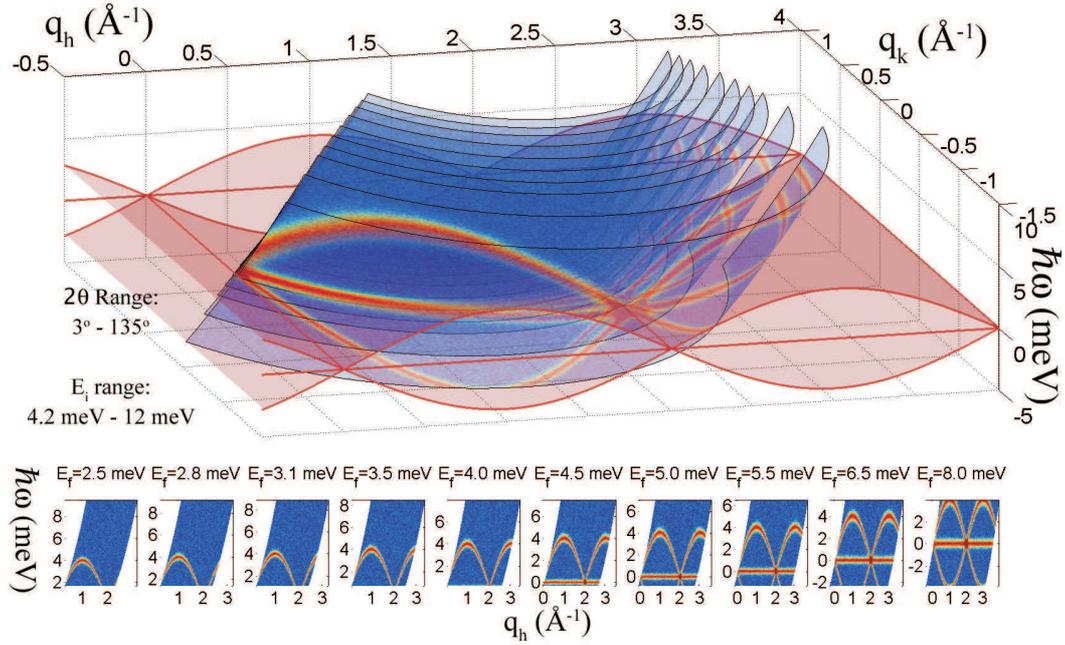}}
\caption{A diagram to represent the measuring capabilities of CAMEA measuring the magnon spin excitation
spectrum of a one dimensional spin system, in a single point scan. The main figure shows the surfaces in
reciprocal space mapped by CAMEA, and the red surface represents the magnon dispersion. Below the main
figure shows the projected excitation spectrums measured by the ten different analysers of CAMEA. No
spinon continuum excitation is included, and dead angles between the analyser sections have been
omitted (equivalent to measuring two points, the second being where the dead angles and active analyser
angles of the secondary spectrometer swap position). }
\label{Scan}       
\end{figure*}

Continuous Angular coverage is obtained by using PSDs that are arranged tangentially. The position along the PSD that the neutron is detected determines the angle at which the neutron was scattered from the sample, and knowing the sample orientation the wavevector of the scattered neutrons is determined. To map out excitations in $(h,\ k,\ \omega)$ of a single crystal a sample rotation scan is performed, in the same way as a scan of a reciprocal plane is produced on a multiplexed spectrometer such as MACS or Flatcone \cite{MACS,flatcone}. In figure \ref{Scan} we show how the magnetic dispersion from a one dimensional spin system can be measured without the need for a sample rotation scan. For Continuous Angular coverage we need to resolve the issue of dead angles between segment wedges of the secondary spectrometer. Continuous Angular coverage can be achieved in two ways, 1) re-position the secondary spectrometer so that the dead angles and active angles swap positions, 2) in a sample rotation scan we can use the detected neutrons observed by different analysers to create a continuous map of the excitations measured with different final neutron energies.

\section{Design}
\label{design}

The specifics of the secondary spectrometer of CAMEA-ESS are shown in figure \ref{Overview}(a) Each segment of the  secondary spectrometer consists of  fifteen wedges of 9$^{\circ}$ width covering a horizontal scattering angle of 3-135$^{\circ}$.   Each wedge consists of 10 PG analysers sat behind each other, with each analyser working at a different final  neutron energy, and three linear PSDs for each analyser. In this way we have ten concentric arcs  of analysers working at ten final neutron energies.

Analysers will be constructed of PG crystals mounted on Si wafers, five or more blades will be used in a vertical focusing Rowland geometry covering  $\sim\pm  1.4^{\circ}$ vertically, to focus the analysed neutrons into three PSD tubes arranged to be parallel to the analysers. An analyser wedge of 9$^{\circ}$ width will consist of 6$^{\circ}$ active analyser, and 3$^{\circ}$ of dead angle for support structures, etc..   As the energy resolution of the secondary spectrometer is limited by distance collimation, the three PSD tubes are at a different scattering angle from a single analyser arc, so each PSD detects a different final neutron energy. This prismatic analysis allows for 30 final energies to be examined by CAMEA at a higher energy resolution\cite{prismatic}.

For the primary spectrometer, CAMEA will be placed on cold neutron moderator. The neutron guide was optimised by simulations using the McStas package and the optimizer package GuideBot\cite{McStas,GuideBot}. A wide range of guide shapes were examined, and the best guide shape identified\cite{mads}.   From the neutron moderator there is a guide feeder to a virtual source at 6.5\,m\cite{bertelsen13}, where a pulse shaping chopper is placed at the closest possible distance to the moderator. To be able to use the full length of the 2.86\,ms neutron pulse of the 14\,Hz ESS, filling the counting window, a pulse shaping chopper at 6.3 m gives a natural length of 165\,m\cite{lefmann-RSI-2013}. 
The neutron guide is then two ellipses separated by a small angular kink in the guide to remove line-of-sight from the neutron source. Elliptical guides reduce neutron losses by reducing the number of reflections required along the guide's length, and the pinch point between the ellipses provide background reduction. A bandwidth chopper will be placed in the kink section of the guide, and the instrument requires three frame overlap choppers, that can be positioned in the first ellipse.

\begin{table*}
\centering
\caption{Specifications for the CAMEA instrument at the ESS.}
\label{tab-1}       
\begin{tabular}{lllllll}
\hline
\bf{Primary Spectrometer}  &   \\\hline\hline
Moderator & Cold   \\
Wavelength Range (Energy Range) & 1\,\AA\  to 8\,\AA\ (81.8\,meV\ to 1.3\,meV)  \\
Bandwidth & 1.7\,\AA  \\
Neutron Guide & 165\,m - Parabolic feeder to double elliptical guide \\
Line-of-Sight Removal & Kink between elliptical guide sections  \\
Number of Choppers & 7, operating from 840\,rpm to 12600\,rpm \\
Beam Divergence & 2.0$^{\circ}$ vertical, 1.5$^{\circ}$ horizontal \\
Divergence Control  & 5 divergence jaws intergrated into the end of the guide \\
Incoming Energy Resolution  & Adjustable from 0.1\,\% to 3\,\% at 5\,meV  \\
Polarizer & Removeable polarizing supermirror s-bender \\\hline
\bf{Sample} &    \\\hline\hline
Maximum Flux on Sample Position & 1.8$\times$10$^{10}$ n/s/cm$^{2}$/1.7\,\AA  \\
Wavector Range at Elastic Position & PG(002) reflections:  0.058\,\AA$^{-1}$  to 3.6\,\AA$^{-1}$  \\
(Including PG(004) reflections)  &  PG(004) reflections:  0.12\,\AA$^{-1}$  to 7.26\,\AA$^{-1}$  \\
Background Count Rate   & < 5$\times$10$^{-5}$ comparted to the elastic signal of vanadium \\
    & (result obtained from prototype testing)  \\
Beam Size at Sample position     & 1.5\,cm $\times$\ 1.5\,cm  \\
Beam Size Resolution Optimization & 0.1\,cm $\times$\ 0.1\,cm -- 1.0\,cm $\times$\ 1.0\,cm   \\
Sample Environment Space & 90\,cm diameter with possible side access \\\hline
\bf{Secondary Spectrometer} &    \\\hline\hline
Collimation & Radial collimation after Sample   \\
 & Cross-talk collimation in secondary spectrometer   \\
Filter  & Removable cooled Be-filter before analyzers   \\
Analyser crystals   & 2\,m$^{2}$ cooled Pyrolytic Graphite (PG)  \\
    & 60" mosaic using (002) and (004) reflections   \\
Detectors     & 2.5\,m$^{2}$ position sensitive  $^{3}$He at 7\,bar  \\
Number of Analyzer Arcs & 10    \\
Number of Analyzer-Detector Segments  & 15 (9.0$^{\circ}$ per segment, 6.0$^{\circ}$ active)   \\
Sample to Analyzer Distances   & 1.00\,m to 1.79\,m   \\
Analyzer to detector Distances    & 0.8\,m to 1.45\,m   \\
Horizontal Angular Coverage &  3$^{\circ}$ to 135$^{\circ}$ \\
Horizontal Angular Resolution & 0.79$^{\circ}$ to 0.46$^{\circ}$  \\
Vertical Angular Coverage  & $\pm$ 1.4$^{\circ}$   \\
Final Neutron Energy  PG(002) &   2.5,\ 2.8,\ 3.1,\ 3.5,\ 4.0,\ 4.5,\ 5.0,\ 5.5,\ 6.5,\ 8.0\,meV \\
Final Neutron Energy Range PG(002) and  PG(004)   & 2.5\,meV to 32\,meV   \\
Secondary Energy Resolution    &  0.77\,\% to 1.3\,\%   \\
Polycrytal  Elastic Wavevector Resolution   & 1.1\,\%\ for $E_f$ = 5.0\,meV  \\
Time Resolution               & 20$\mu$s   \\
Neutron Polarization & Polarizing supermirrors   \\\hline
\end{tabular}
\end{table*}

A prototype of the secondary spectrometer of CAMEA has been built, and tested on the MARS spectrometer at the SINQ neutron source, Paul Scherrer Institut, Switzerland. The results of this testing  validates the CAMEA design and  will be reported in detail elsewhere\cite{Marci}. In addition to this, analytical calculations of CAMEA have been performed\cite{calculations}.

For any spectrometer background reduction is critical, and requires a clear strategy to achieve. A set of beam definition jaws at the end of the guide section will define the beam divergence that reaches the sample position, as has been implemented on WISH at the ISIS facility\cite{WISH}. The divergence jaws are followed by diaphragm to define the beam size at the sample position. Any neutrons that pass straight through the sample will be directed along a get lost tube to the beam stop. The secondary spectrometer will either be in an Ar atmosphere or under vacuum to remove background from air scattering of neutrons. Line-of-sight between the PSDs and the sample position will be shielded by neutron absorbing materials. Radial collimation between the sample and analysers, and cross talk collimation inside the secondary spectrometer are foreseen.  A removable Be-filter can be placed in the scattered neutron beam to remove high energy neutrons, although this restricts CAMEA to working with the first seven analysers. The effectiveness of our background reduction strategy has been confirmed in our prototype testing of CAMEA, where a the background count rate is 5$\times$10$^{-5}$ times that from  the incoherent scattering from a vanadium sample.


\begin{figure}
\includegraphics[width=7.5cm,clip]{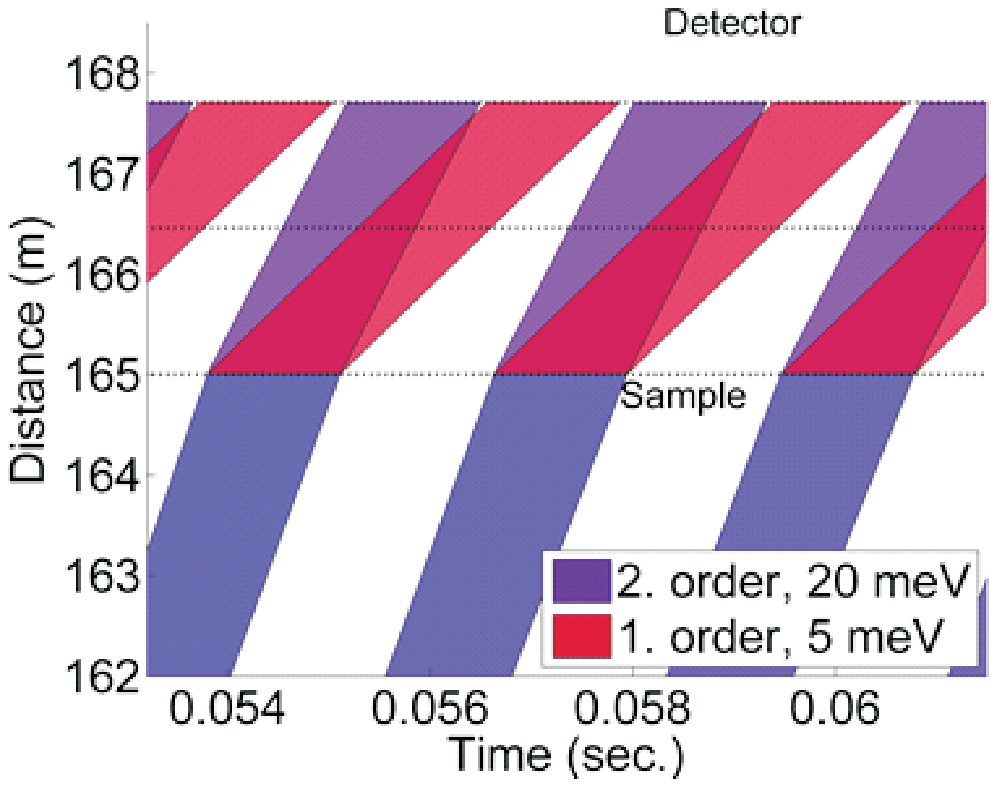}
\caption{Time-of-flight diagram of the order sorting chopper and the 7th  PG(002) $E_f = 5$\,meV analyser. At 162\,m the chopper divides the pulse into $\sim 26$ pulses per source pulse, at 3\,m from the detector the neutrons hit the sample and scatters. The time distance diagram clearly shows how the neutrons that scatter of the analyser at 166.5 m by the first and second order reflections are detected over unique time windows at the detector. The time gaps between the signals at the detector are necessary due to a small time broadening of the signals that is mainly due to the choppers open and closing time.}
\label{ordersort}
\end{figure}

The energy range can be expanded to measure excitations of thermal energies on CAMEA using the order sorting chopper. If thermal neutrons are scattered off the sample, the scattered neutrons can be analysed by the PG(002) or PG(004) analyser reflections with $E_f (PG(004))$ = 4$\times E_f (PG(002))$, and we cannot directly differentiate between them. In figure \ref{ordersort}  we show a time-distance plot for the order sorting chopper. The order sorting chopper for CAMEA consists of two counter rotating disc choppers that run at 180 Hz with two symmetric openings, effectively 360\,Hz a non-integer multiple of the the source's repetition rate. The time distance plot of Fig. \ref{ordersort} for the $E_f (PG(002)) = 5$\,meV shows that neutrons that arrive at the detector at a given time are uniquely determined from the PG(002) or PG(004) analyser reflection. Using the order sorting chopper CAMEA has analysers working from  $E_f = 2.5$\,meV to 32\,meV. The order sorting chopper however reduces the incident neutron intensity by $59\,\%$ when placed 3\,m from the sample position. The greater the distance for the order sorting chopper to the sample position, the greater the reduction in incident neutrons, 3\,m represents a safe distance from stray fields from future >20\,T cryomagnet possibilities.

\section{Performance}
\label{performance}

In table one we outline specifications of the CAMEA spectrometer. Ten analysers represents a compromise on coverage, the cost increase of the analyser for increasing sample to analyser distance, and the reduced performance due to transmission losses. The CAMEA analyser resolution is limited by distance collimation, not mosaic quality, so 60' mosaic PG crystals can be used to improve count rates without reducing the energy resolution. When the primary and secondary resolutions are matched, CAMEA will achieve a higher energy resolution than than triple-axis spectrometers, whose energy resolution is limited by mosaic quality.

If the full ESS pulse width is used the primary spectrometer energy resolution is 4\,$\%$, with 1.8$\times$10$^{10}$ neutrons s$^{-1}$  cm$^{-2}$ for a 1.7\,$\AA$ bandwidth centred at 3\,$\AA$\cite{mads}. We consider the simulated performance of ThALES the upgraded IN14 at the ILL, which to the best of our knowledge will be the highest flux cold TAS in the world with a monochromatic flux maximum of 3.5$\times$10$^{8}$\cite{THALES}, a factor of 50  lower than the maximum polychromatic flux of CAMEA. The energy resolution of the primary spectrometer of CAMEA can be improved to  0.8\% to match the secondary spectrometer, with the primary resolution being directly proportional to the pulse width from the pulse shaping chopper. For example the flux for 2\% total energy resolution on CAMEA is 0.9$\times$10$^{10}$.

Within $\pm  1.4^{\circ}$ vertical range of CAMEA, and taking into account transmission rates, we estimate the solid angle gain of CAMEA over Flatcone to be 36\cite{flatcone}, or 23 over MACS at NIST\cite{MACS}. Comparing CAMEA's flux gain for 2\% energy resolution, and the increased solid angle coverage gives a gain factor of $\sim$ 900 compared to  ThALES using Flatcone, provided all detected signal is of use in both cases. With this gain factor CAMEA will enable inelastic neutron scattering on samples of $\bf{1\,mm^{3}}$ and smaller as a routine measurement. Neutron simulations were also used to compare CAMEA to a 150\,m long cold direct ToF spectrometer that uses Repetition Rate Multiplication, with  a large vertical  coverage of $\pm 30^{\circ}$, at the ESS. These simulations indicated within the horizontal plane CAMEA has a 22 times higher count rate than the direct ToF\cite{birk}. If we consider the total counts of the direct ToF ESS spectrometer including a large vertical angular coverage, CAMEA has a slightly higher count rate.

\begin{table*}
\centering
\caption{Demand for several European based cold TAS and indirect geometry spectrometers that measure magnetic excitations in single crystals, and the
demand for extreme environments conditions on these instruments. The overload of an instrument is defined as
the number of days applied for experiments divided by the total number of days available to perform experiments.}
\label{tab-2}       
\begin{tabular}{llllll}
\hline
Instrument  & Overload & Magnetic Fields & Pressure  & $\leq 1$,\ K  & Polarized neutrons
 \\
 (Instute/Neutron Source) &  & (\% of proposals ) & (\%) & (\%) & (\%)  \\\hline
RITA-II, TASP (PSI) & 2.5 & 34 & 4 & 19 & N/a  \\
PANDA (FRM-II) & 2.7 & 30 & 5 &  20 & N/a  \\
IN14 (ILL) & 2.5 & 30-40 & $<$ 5 & 60 & 20-25    \\
IN12 (JCNS@ILL) & 2.6 & 24 & 0 & 28 & 10   \\
Osiris (ISIS) & 2 & 40 & 0 & 40 & Planned   \\
FLEX (HZB) & 1.5 & 56 & 0 & 20 & Commissioning \\\hline
\end{tabular}
\end{table*}

Polarized inelastic neutron scattering will be available for CAMEA from the beginning. The incident neutron beam will be polarized by a polarizing supermirror S-bender, inserted near the end of the guide by a guide changer, a setup successfully used on instruments such as FLEXX\cite{FLEXX}. Polarization analysis of the scattered beam will be performed by a wide angle polarized supermirror analyser replacing the Be filter, which is equivalent to the setup used on the D7 instrument\cite{D7}. We chose a polarized supermirror analyser over a wide angle He-3 polarization cell, to enable the use of cryomagnets which produce stray fields, and to keep a large sample environment space available with the polarized option.

\section{Scienctific Demand for Extreme Conditions}
\label{demand}

The CAMEA geometry is especially suitable for inelastic neutron scattering in applied magnetic fields, and under extreme pressure. In table two we outline the results of a survey into the use of extreme environments on cold inelastic neutron spectrometers in Europe. Typically there are overload factors of 2.5 for these instruments, with $\sim 33\%$ requesting the use of a cryomagnet.  The lack of demand for use of high pressure for neutron spectroscopy is likely due to the highly restrictive sample volume of pressures cells, and the present need for large single crystals for inelastic studies. In September 2012 attendees of an ESS Science Symposium on Strongly Correlated Electron Systems were asked to name three instruments you would like to have at the ESS, and a spectrometer for   extreme conditions was one of the three instruments\cite{SCESESSmeeting}. There is a significant scientific demand for spectroscopy under extreme conditions in Europe, which the ESS can accommodate through CAMEA.

\section{Scienctific Capabilities of CAMEA}
\label{ability}

The massive gain factor of CAMEA ESS has the potential to enable scientific discoveries in several fields of research, a few of which are discussed below.

The present user community for using extreme environments in inelastic neutron scattering is based in the magnetic scattering community, which includes magnetic materials, strongly correlated electron systems, superconductors, quantum magnets, etc..  CAMEA offers this community high counting rates to study weak excitations to a level of accuracy
that present instrumentation cannot reach, bridging the gap between the accuracy theoretical calculations can reach and present inelastic neutron scattering. Alternatively rapid mapping of excitations will be enabled for parametric studies of dynamics across critical transitions, providing a unique tool to study wavevector and energy evolution across transitions. The large gain factor of the instrument and the low background count rate will unlock the ability to study magnet materials under extreme pressure.


At present the sample size required for inelastic neutron scattering is prohibitive, limiting the technique
to crystals grown by techniques such as floating zone mirror light furnaces. The ability to study samples of less than $\bf{1\,mm^{3}}$ opens up the possibility for sample growth for neutron scattering in both material discovery and soft matter. For example it will be possible to study materials grown by high-pressure synthesis (which is how the highest Tc iron-based superconductors were first crystalized) and hydrothermal synthesis (which is how the best known realization of a kagome quantum magnet is synthesized) CAMEA will enable neutron scattering to be be part of the iterative process to discover new materials classes. This will lead to input from inelastic neutron scattering immediately after materials are discovered, or directly lead to discovery of materials. At present a large amount of experimental and theoretical work is wasted due to incorrect assumptions made about the spin and lattice interactions in materials, inelastic neutron scattering unambiguously resolves these issues.

CAMEA has the potential to open up the application of neutron spectroscopy in new fields of reasearch including biosphysical studies of collective dynamics in membranes. In membranes collective dynamics are believed to drive transport of molecules, pore opening, membrane fusions and protein-protein interactions\cite{Rheinstadter12}, which can be determined by inelastic neutron scattering. At present studies of collective dynamics in membranes by neutrons is restricted to model systems which can be prepared in large multi-layer stackes, the small sample capability of CAMEA will enable studies of the actual membranes of interest.

There exist a great hitherto unaccommodated interest to study lattice dynamics in simple materials under
extreme pressure, and for geo- and planetary science related studies such as hydrogen diffusion in materials of
the Earth’s upper mantle. CAMEA is ideally suited for both of these purposes. Despite the fact that water is vital for life on Earth we have little knowledge on the extent of the water cycle in the Earth’s mantle. Estimates on the water in the mantle wildly vary from ten percent to two and a half times the water on the Earth’s surface \cite{water}. The uptake of water into the material of the Earth’s mantle greatly influences the properties of the materials, which has consequences for flow of material and sound velocities in the mantle, studying these materials has the potential to provide great insight into plate-tectonics and seismic activity\cite{water,natwater}. To study the effects of hydrogen on the different phases of the material of the Earth's mantle requires performing neutron scattering at pressure up to 30\,GPa for temperatures of the order of 2000\,K. The experimental conditions imply a pressure cell  with a sample volume $< 5$\,mm$^{3}$\cite{Klotz}, and an instrument resolution of a cold TAS is required, well within CAMEA's capability\cite{bove}.

\begin{table}
\centering
\caption{Desirable sample environments for CAMEA, within predicted technical developments. }
\label{tab-3}       
\begin{tabular}{lllllll}
\hline
Sample Environment  & Performance   \\\hline
Low Temperatue & Dilution to <100\,mK   \\
Magnetic fields  & Vertical >20\,T  \\\hline
Pressure & 30 GPa with 5\,mm$^{3}$ sample,    \\
 & T = 3 -- 2000\,K  \\
& 10 GPa with 50\,mm$^{3}$ sample, \\
 & T = 0.1 -- 1800\,K    \\\hline
Magnetic Field  & >10\,T with upto 10 GPa \\
and Pressure & T =  0.1 -- 350\,K  \\\hline
\end{tabular}
\end{table}

In table 3 we outline some of the desirable sample environment for CAMEA to perform these experiments.

For time dependent studies the time resolution of CAMEA is only from the secondary spectrometer, 
and of the order of 20$\mu$s. An analyser arc of CAMEA measures one excitation energy for a time of the order of the 2.86\,ms source pulse width with 20$\mu$s resolution, that is the time dependence of excitations at ten different energies are simultaneously measured. This capability of CAMEA opens up experimental possibilities in inelastic neutron scattering for example in soft matter stimulated out of equilibrium by pump-probe techniques, or studying excitations in pulsed high magnetic fields beyond 30\,T.

\section{Conclusion}

CAMEA provides an evolution in cold neutron indirect spectroscopy by performing analysis of the energy of the
scattered neutrons at 10 final energies, that can be increased to 30 energies by prismatic analysis. This instrument has been designed through simulations, with validation
of the simulations of the secondary spectrometer achieved by prototype testing on the MARS spectrometer at
the SINQ neutron source of  the Paul scherrer Insitut. Compared to present cold multiplexed TAS CAMEA has three orders of magnitude gain.  At the ESS the in plane count rate of CAMEA is over an order of magnitude higher than cold direct geometry ToF spectrometers, and is equivalent to the total count rate of a cold direct ToF spectrometer.  CAMEA therefore enables inelastic neutron scattering on samples of less than 1\,mm$^{3}$ as a routine measurement, enabling experiments in fields of research such material discovery , soft matter, and extreme pressure studies in magnetism, and geoscience. Finally we note that the secondary spectrometer of CAMEA can be implemented as a multiplexing option for a TAS instrument, that could perform within a factor of 100 of CAMEA at the ESS. A CAMEA TAS is being built for the the RITA-II spectrometer at SINQ neutron source, P.S.I., Switzerland.

The work presented here is part of the European Spallation Source Design Update Programs of Switzerland and Denmark.

\end{document}